\documentclass[10pt,aps,twocolumn,superscriptaddress,longbibliography]{revtex4-1}
\usepackage[resetlabels]{multibib}
\usepackage{graphicx, amsmath, verbatim, dsfont, amsfonts, color}
\usepackage{bbm, bm, latexsym, amssymb}
\usepackage[us]{datetime}
\usepackage{color}
\usepackage{subcaption}
\usepackage{tikz,xcolor,hyperref}
\usepackage[T1]{fontenc}
\definecolor{lime}{HTML}{A6CE39}
\DeclareRobustCommand{\orcidicon}{%
	\begin{tikzpicture}
	\draw[lime, fill=lime] (0,0)
	circle [radius=0.16]
	node[white] {{\fontfamily{qag}\selectfont \tiny ID}};
	\draw[white, fill=white] (-0.0625,0.095)
	circle [radius=0.007];
	\end{tikzpicture}
	\hspace{-2mm}
}

\foreach \x in {A, ..., Z}{%
	\expandafter\xdef\csname orcid\x\endcsname{\noexpand\href{https://orcid.org/\csname orcidauthor\x\endcsname}{\noexpand\orcidicon}}
}

\begin{document}

\title{Superexchange dominates in magnetic topological insulators}

\author{Cezary \'{S}liwa}\email{sliwa@ifpan.edu.pl}
\affiliation{Institute of Physics, Polish Academy of Sciences,
Aleja \ Lotnikow 32/46, PL-02668 Warsaw, Poland}

\author{Carmine Autieri\orcidA}
\affiliation{International Research Centre MagTop, Institute of Physics, Polish Academy of Sciences, Aleja Lotnikow 32/46, PL-02668 Warsaw, Poland}
\affiliation{Consiglio Nazionale delle Ricerche CNR-SPIN, UOS Salerno, I-84084 Fisciano (Salerno), Italy}

\author{Jacek A. Majewski}
\affiliation{Institute of Theoretical Physics, Faculty of Physics, University of Warsaw,
ul. Pasteura 5, PL-02093 Warsaw, Poland}

\author{Tomasz Dietl\orcidB}\email{dietl@MagTop.ifpan.edu.pl}
\affiliation{International Research Centre MagTop, Institute of Physics, Polish Academy of Sciences, Aleja Lotnikow 32/46, PL-02668 Warsaw, Poland}
\affiliation{WPI-Advanced Institute for Materials Research, Tohoku University, Sendai 980-8577, Japan}


\begin{abstract}
It has been suggested that the enlarged spin susceptibility in topological insulators, described by interband Van Vleck's formalism, accounts for the ferromagnetism of bismuth-antimony topological chalcogenides doped with transition metal impurities. In contrast, earlier studies of HgTe and related topological systems pointed out that the interband analog of the Ruderman-Kittel-Kasuya-Yosida interaction (the Bloembergen-Rowland mechanism) leads to antiferromagnetic coupling between pairs of localized spins. Here, we critically revisit these two approaches, show their shortcomings, and elucidate why the magnitude of the interband contribution is small even in topological systems. From the proposed theoretical approach and our computational studies of magnetism in Mn-doped HgTe and CdTe, we conclude that in the absence of band carriers, the superexchange dominates, and its sign depends on the coordination and charge state of magnetic impurities rather than on the topological class of the host material.
\end{abstract}

\maketitle

In the traditional approach to localized magnetism in solids, one considers pairwise exchange interactions between spins $J_{ij}$ comprising the Ruderman-Kittel-Kasuya-Yosida (RKKY) coupling brought about by band carriers and the Anderson-Goodenough-Kanamori superexchange mediated mainly by anion orbitals \cite{White:2007_B}.  However, it has been demonstrated that in the case of p-type dilute magnetic semiconductors (DMSs), the Zener model \cite{Zener:1951_PRa} is remarkably versatile,  in which the local magnetization $\mathbf{M}(\mathbf{r})$ plays a role of a continuous order parameter. This approach has allowed understanding the physics of bound magnetic polarons \cite{Dietl:1982_PRL,Dietl:2015_PRB} and ferromagnetism of p-type DMSs \cite{Dietl:2000_S}, and subsequently describing quantitatively a wealth of micromagnetic properties and spintronic functionalities of (Ga,Mn)As and related ferromagnets \cite{Dietl:2014_RMP,Jungwirth:2014_RMP}. Notably, the equivalence between the RKKY and  Zener models was established within the mean-field approximation (MFA) \cite{Dietl:1997_PRB}.

Ferromagnetic topological insulators \cite{Ke:2018_ARCMP,Tokura:2019_NRP}, such as (Bi,Sb,Cr)$_2$Te$_3$, have made possible the experimental realization of the quantum anomalous Hall effect \cite{Chang:2013_S}, the axion insulator \cite{Xiao:2018_PRL}, efficient magnetization reversal by spin currents \cite{Fan:2014_NM}, and the much disputed chiral Majorana fermions \cite{He:2017_S,Kayyalha:2020_S}. Interestingly, the appearance of ferromagnetism in these systems is also attributed to their topological character, as the inverted band structure enhances the interband spin susceptibility leading to carrier-independent spin-spin coupling \cite{Yu:2010_S}, referred to as the Van Vleck magnetism \cite{Ke:2018_ARCMP,Tokura:2019_NRP,Yu:2010_S}. That appears surprising, however,  as early studies of spin-spin coupling mediated by an interband analog of the RKKY interaction (the Bloembergen-Rowland (BR) mechanism \cite{Bloembergen:1955_PR}) found predominately {\em antiferromagnetism} in topological Mn-doped zero-gap topological HgTe \cite{Bastard:1979_PRB,Lewiner:1980_JPC,Lee:1988_PRB}.

In this Letter, we resolve this puzzle by demonstrating that the mean-field Zener-Van Vleck model fails in the case of magnetism associated with interband bulk excitations in insulators. Furthermore, by making use of the recent progress in the theory of the indirect exchange interaction \cite{Sliwa:2018_PRB} and in the quantitative description of exchange splitting of bands in the whole Brillouin zone (BZ) \cite{Autieri:2021_PRB}, we determine various contributions to spin-spin coupling in non-topological CdTe and topological HgTe doped with Mn ions (see Supplemental Material at the manuscript end for an introduction to the topic). We find that the superexchange dominates not only in (Cd,Mn)Te \cite{Larson:1988_PRB,Savoyant:2014_PRB}, but also in topological (Hg,Mn)Te. We also show that the conclusion about the predominant role of the superexchange substantiates the experimental results on (Cd,Hg,Mn)Te \cite{Spalek:1986_PRB,Lewicki:1988_PRB} and explains hitherto challenging chemical trends in the magnetic properties of V, Cr, Mn, and Fe-doped tetradymite topological insulators observed experimentally \cite{Ke:2018_ARCMP,Tokura:2019_NRP,Satake:2020_PRM} and found in {\em ab initio} studies \cite{Vergniory:2014_PRB,Kim:2017_PRB,Peixoto:2020_QM}.

{\em RKKY-BR vs. Zener-Van Vleck models.} For concreteness, we consider $xN$ randomly distributed Mn spins $S=5/2$  in zero-gap Hg$_{1-x}$Mn$_x$Te in which both the conduction and valence bands are of $\Gamma_8$ symmetry at the BZ center. In the high-temperature expansion of the partition function for the pairwise interactions \cite{Spalek:1986_PRB}, the contribution of the RKKY-BR term to the Curie-Weiss temperature (equal to spin ordering temperature $T_{\text{c}}$ within MFA) assumes the form \cite{Bastard:1979_PRB,Lee:1988_PRB,Dietl:1994_B},
\begin{widetext}
\begin{equation}
\Theta_{\text{CW}} = \frac{xS(S+1)}{3N{\mathcal{V}}}\sum_{i \ne j,\mathbf{q}}\exp[\mbox{i}\mathbf{q}\cdot(\mathbf{R}_i -\mathbf{R}_j)]
\sum_{\mathbf{k}, n, n', \sigma, \sigma'}\frac{2|\langle u_{n, \mathbf{k}, \sigma}|\beta s_z|
u_{n', \mathbf{k + q}, \sigma'}\rangle|^2}{{\mathcal{V}}(E_{n',\mathbf{k+q}}- E_{n,\mathbf{k}})}f_{n,\mathbf{k}}(1-f_{n',\mathbf{k+q}}),
\label{eq:Theta}
\end{equation}
\end{widetext}
where the Boltzmann constant $k_{\text{B}} =1$; ${\cal{V}}$ is the crystal volume,  and $\mathbf{k}, \mathbf{k+q} \in \text{BZ}$. We see that if the $p$-$d$ exchange integral $\beta$ were $\mathbf{k}$-independent, the summation over $\mathbf{k}$ would provide the spin susceptibility $\tilde{\chi}(\mathbf{q})$ of the $\Gamma_8$ bands $n$ and $n'$, as defined in Refs.\,\onlinecite{Dietl:2001_PRB} and \onlinecite{Ferrand:2001_PRB}. Furthermore, if the contribution of the self-interaction energy were small compared to interaction energies for $i \ne j$,  the term $i =j$ could be included in Eq.\,\ref{eq:Theta}, transferring the sum over the cation positions $\mathbf{R}_i$ and $\mathbf{R}_j$ into the structure factor that is non-zero for $\mathbf{q}=0$ only. This is the case of the long-range RKKY coupling, for which the sum over $i,j$ can be approximated by $N^2\beta^2\tilde{\chi}(0)$, as presumed within the Zener-Van Vleck model \cite{Dietl:2001_PRB,Ferrand:2001_PRB,Yu:2010_S}.  However,  in the case of the BR mechanism, the decay of the interaction with the inter-spin distance is faster \cite{Bloembergen:1955_PR,Bastard:1979_PRB,Lewiner:1980_JPC,Lee:1988_PRB} and, therefore, $\tilde{\chi}(q)$ beyond $q =0$ determines the sign and magnitude of $\Theta_{\text{CW}}$ and $T_{\text{C}}$. In conclusion, atomistic computations of pair exchange energies $J_{ij}$ are necessary in order to  meaningfully evaluate the role of the interband contribution \cite{Ginter:1979_pssb}.

{\em Theoretical methodology.} We consider exchange interactions between Anderson magnetic impurities occupying cation substitutional positions in considered semiconductor compounds, whose band structures are described within the empirical tight-binding approximation taking into account spin-orbit interactions. This approach \cite{Blinowski:1994_APPA,Blinowski:1996_PRB,Simserides:2014_EPJ} was successfully applied to elucidate the nature of ferromagnetism in (Ga,Mn)N \cite{Sawicki:2012_PRB,Stefanowicz:2013_PRB} and has recently been generalized by us to simultaneously take into account various contributions to the spin pair exchange energy \cite{Sliwa:2018_PRB}, including the BR interband term. Assuming time-reversal symmetry (no spontaneous magnetization) and within the fourth order perturbation theory in the $p$-$d$ hybridization energy $V_{\text{hyb}}$ between band states $E_{\mathbf{k},n}$ and Mn $d$ orbitals residing at $E_d = E(d^5) - E(d^4)$  and $E_d + U = E(d^6) - E(d^5)$, the spin Hamiltonian is,
%
%
%
\begin{equation}
  \hat {\cal{H}}_{\mathrm{eff}}^{(4)} = - \sum_{i \ne j} J^{(4)}_{ij,\alpha\beta} \hat S_{i,\alpha} \hat S_{j,\beta},
\end{equation}
where the tensor of exchange integrals (parameters) for spins at sites $(i, j)$ can be written as a double integral
over the BZ [$k \equiv (\mathbf{k}, n)$]:
\begin{equation}
  J^{(4)}_{ij,\alpha\beta} = -\frac{1}{2(2S)^2} \sum_{m, m'} \sum_{k, k'} A^{(4)}_{k k'} W_{i \alpha, k' k, m} W_{j \beta, k k', m'},
\end{equation}
where $m$ labels the $d$ orbitals and
\begin{eqnarray}
\lefteqn{ W_{i \alpha, k' k, m} = } \nonumber \\
&=&   \sum_{a, b = \uparrow, \downarrow}\left< k' \middle| V_{\mathrm{hyb}}^{\dagger} \middle| d_i m a \right>
       \left< a \middle| \sigma_\alpha \middle| b \right> \left< d_i m b \middle| V_{\mathrm{hyb}} \middle| k \right>,
\end{eqnarray}
where $a$ and $b$ are spin directions and $\sigma_\alpha$ the Pauli matrices.
For an insulator in a zero-temperature approximation and using the notation,
\begin{eqnarray}
  w_k = \frac{1}{E_d + U - E_k}; \quad
  w'_k = \frac{1}{E_d - E_k}, \label{eq:wk}\\
  w_{k'} = \frac{1}{E_d + U - E_{k'}}; \quad
  w'_{k'} = \frac{1}{E_d - E_{k'}},
 \end{eqnarray}
one can write $A^{(4)}_{kk'}$ in terms of the Heaviside step function $\Theta$ as,
\begin{widetext}
\begin{eqnarray}
    A^{(4)}_{kk'} \approx & \Theta(E_F-E_k) \Theta(E_F-E_{k'}) w_k w_{k'}  \left( w_k + w_{k'} \right)
  + \Theta(E_F-E_k) \Theta(E_{k'}-E_F) \frac{\left( w_k - w'_{k'} \right)^2}{E_k - E_{k'}} \nonumber \\
   & {} + \Theta(E_k-E_F) \Theta(E_F-E_{k'}) \frac{\left( w_{k'} - w'_k \right)^2}{E_{k'} - E_k}
  - \Theta(E_k-E_F) \Theta(E_{k'}-E_F) w'_k w'_{k'} \left( w'_k + w'_{k'} \right) \nonumber \\
 & + \frac{2}{U} \Biggl[ \Theta(E_F-E_k) w_k
             + \Theta(E_k-E_F) w'_k \Biggr] \Biggl[ \Theta(E_F-E_{k'}) w_{k'}
             + \Theta(E_{k'}-E_F) w'_{k'} \Biggr],
  \label{eq: a4}
\end{eqnarray}
\end{widetext}
where we assume that the Fermi energy $E_F$ lies in the range $E_d < E_F < E_d + U$.
Average values of the exchange integrals and $\Theta_0 =\Theta_{\text{CW}}/x$ are obtained by tracing the tensors,
\begin{equation}
J^{(4)}_{ij} = \frac{1}{3} \sum_\alpha J^{(4)}_{ij,\alpha\alpha};\,\,\, \Theta_{0}= \frac{2}{3}S(S + 1) \sum_{i \ge 1} z_i J^{(4)}_{0i},
\label{eq:Theta_0}
\end{equation}
where $z_i$ is a number of cation sites in the consecutive  coordination spheres $i \ge 1$.

\begin{figure}[tb]
\includegraphics[width=1.0\columnwidth]{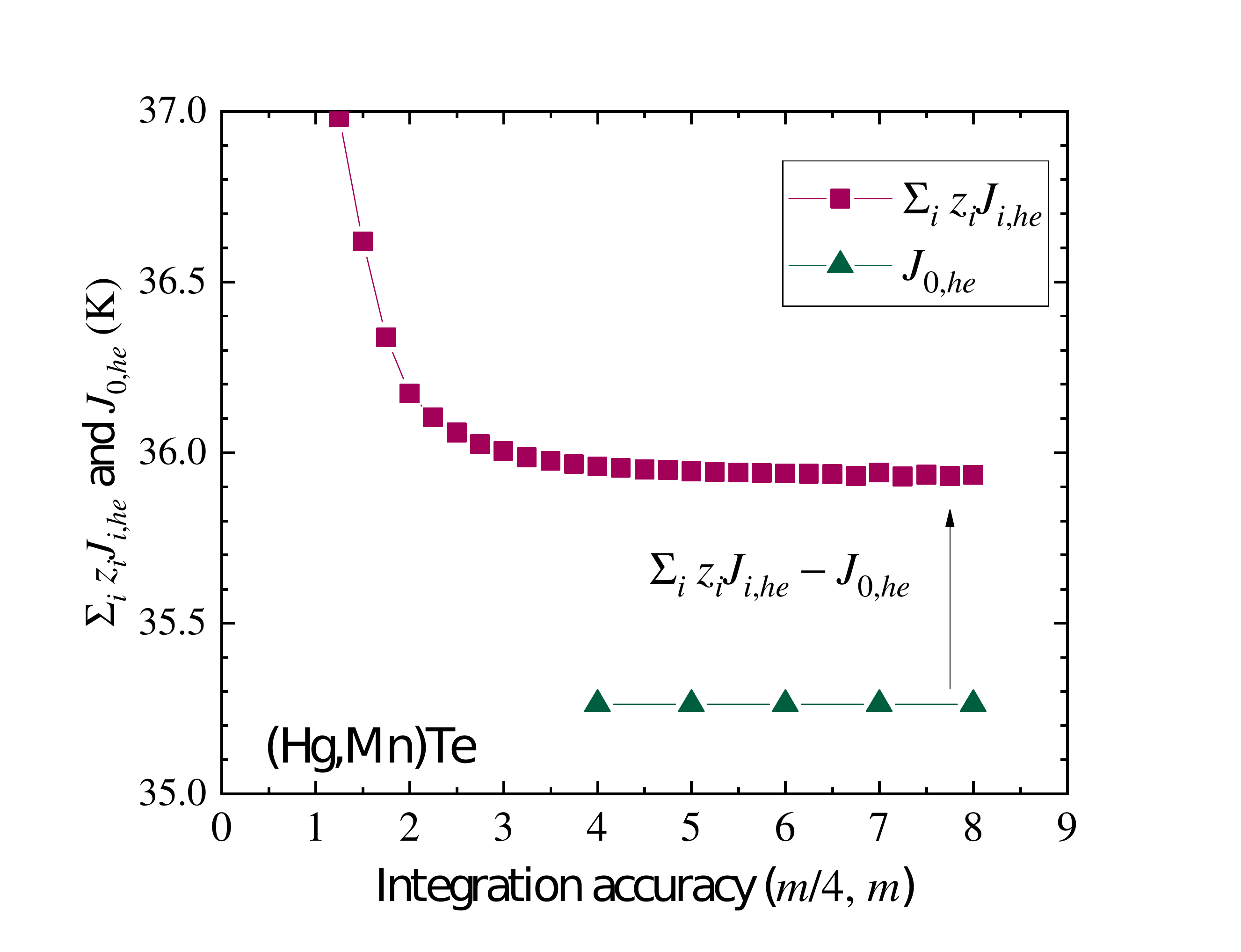}
\caption{Convergence of the interband $he$ term for Mn pairs in HgTe
with respect to the number $m$ of $\vartheta$-points in each space direction in the
trapezoids quadrature: squares --- the sum of exchange integrals
including $J_0$ calculated for the $4\times4\times4$ supercell as a single BZ integral (abscissa provides $m/4$);  triangles ---
the value of $J_0$ to be subtracted in order to obtain $\Theta_{\text{CW}}$ (abscissa provides $m$).}
\label{fig: cvg}
\end{figure}

We represent $A^{(4)}_{kk'}$ as a sum of three contributions \cite{Larson:1988_PRB,Kacman:2001_SST}: the superexchange (or $hh$) term
includes contributions proportional to $\Theta(E_F-E_k) \Theta(E_F-E_{k'})$, the two-electron (or $ee$) term
includes those proportional to $\Theta(E_k-E_F) \Theta(E_{k'}-E_F)$, and the electron-hole ($he$) includes
those proportional to $\Theta(E_F-E_k) \Theta(E_{k'}-E_F)$ or $\Theta(E_k-E_F) \Theta(E_F-E_{k'})$.
Such a decomposition leads to the analogous decomposition of $J^{(4)}_{ij}$ and $\Theta_{0} =
 \Theta_{hh} + \Theta_{ee} + \Theta_{he}$.

We use  a 16-orbital $sp^3$ tight-binding model of the band structure together with the parameter values obtained recently by us employing a modified GGA + U {\em ab initio} approach with $U_{\text{Mn}} = 5$\,eV. These parameters are presented in Tables~II and III of Ref.~\onlinecite{Autieri:2021_PRB}. In particular, $E_d$ in our Eqs.\,\ref{eq:wk}-\ref{eq: a4} is the mean value of $E_{eg\uparrow}$  and $E_{t2g\uparrow}$, whereas $E_d+U$ is the mean values of $E_{eg\downarrow}$  and $E_{t2g\downarrow}$. Finally, matrix elements of $V_{\text{hyb}}$ are spin-averaged values of $V_{sd}\sigma$, $V_{pd}\sigma$, and $V_{pd}\pi$ \cite{Autieri:2021_PRB}. Extensive magnetooptical data collected for Cd$_{1-x}$Mn$_x$Te and Hg$_{1-x}$Mn$_x$Te at the $\Gamma$
and $L$ points of the BZ served to benchmark the model \cite{Autieri:2021_PRB}. Importantly, our tight-binding model reconfirms for these compounds stronger hybridization of $t_{2g}$ orbitals with the band states, compared to the $e_g$ case, which is crucial for the sign and magnitude of the interaction between localized spins.

In topological materials, the most interesting is the $he$ term. It appears whenever transitions between
the fully occupied valence bands and the empty conduction bands are symmetry-allowed (the BR mechanism),
or when there is a non-zero density of states at the Fermi level (the RKKY mechanism).
This term features an energetic denominator $E_k - E_{k'}$. In insulators, the latter
is guaranteed to be non-zero by the Heaviside-$\Theta$ prefactors (it is understood that
each term vanishes whenever the zero Heaviside--$\Theta$ prefactor does, even despite singular denominators).
However, in a semimetal, the denominator is singular at the Fermi level (i.e., when either $E_k \to E_F^{+}$
and $E_{k'} \to E_F^{-}$, or vice versa). In undoped (or isoelectronically doped) HgTe,
this happens at the $\Gamma$ point of the BZ. For this reason, it
has been essential to elaborate a special integration method.
The Supplemental Material at the manuscript end presents issues associated with
$k$ and $k'$ integrations in simple models \cite{Bastard:1979_PRB,Lewiner:1980_JPC,Lee:1988_PRB}
and  a comparison of the second-order perturbation theory in the $p$-$d$ exchange
integral  compared to the  fourth-order perturbation theory in the
hybridization matrix element  $V_{\text{hyb}}$ employed here.

{\em{Brillouin-zone integration}}. Although our goal is to find the exchange integrals $J^{(4)}_{ij}$ in the limit of an infinitely large system,
it is typical in numerical calculations to replace the BZ integration by a summation over a discrete set of
$\mathbf k$-points in the BZ.
As pointed out in Ref.\ \onlinecite{Froyen:1989_PRB}, such discrete $\mathbf k$-point mesh may be defined
through the introduction of the superlattice vectors $\{ \mathbf g_i \}$, being the linear
combinations of primitive crystal translations $\{ \mathbf a_i \}$,
\begin{equation}
  \mathbf g_i = \sum_{j = 1}^3 \mathbf a_j M_{ji}.
\end{equation}
In cubic systems the three $\mathbf g_i$ can be taken as vectors along the three Cartesian axes
with length $L a$. This defines the equidistant $\mathbf k$-point mesh $\mathbf \kappa_{\{m\}} =
\frac{2 \pi}{L a} (m_x, m_y, m_z)$ of $L^3$ grid points. In addition, the grid points can be
shifted.

However, the expressions for the $J_{ij}$ tensor components involve double integration over the BZ.
Here, we present an efficient method to deal with the double BZ integration facilitated
by the specifics of the integrand. In particular, this method allows for accurate treatment
in the $he$ contribution for $J_{ij}$ in the case of zero-gap systems.
Indeed, the product $W_{i \alpha, k' k, m} W_{j \beta, k k', m'}$
includes a phase factor $\exp[\mbox{i}(\kappa - \kappa') (R_i - R_j)]$, and the summation
over the images of $R_i$, $R_i + L a_{\text{lat}} (n_x, n_y, n_z)$ yields (by the principle of the Poisson summation)
a set of Dirac deltas at $\kappa - \kappa' = \frac{2 \pi}{L a_{\text{lat}}} (m_x, m_y, m_z)$,
\begin{widetext}
\begin{eqnarray}
  \lefteqn{\sum_{n_x, n_y, n_z}\exp\left\{\mbox{i}(\kappa - \kappa') [R_i - R_j + L a_{\text{lat}} (n_x, n_y, n_z)]\right\} = } \nonumber \\
  & = & \left(\frac{2\pi}{L a_{\text{lat}}}\right)^3 \exp\left[\mbox{i} (\kappa - \kappa') (R_i - R_j)\right]
    \sum_{m_x, m_y, m_z} \delta\left(\kappa - \kappa' + \frac{2\pi}{L a_{\text{lat}}} (m_x, m_y, m_z)\right), \quad
  \label{eq: ps}
\end{eqnarray}
where the delta suppresses only one integration.
To handle this issue, we first sum over a shifted grid of equidistantly spaced $k$-points,
\begin{equation}
  \kappa = \frac{2 \pi}{L a_{\text{lat}}} (m_x + \frac{\vartheta_x}{2\pi}, m_y + \frac{\vartheta_y}{2\pi}, m_z + \frac{\vartheta_z}{2\pi}); \,\,\,\,
  \kappa' = \frac{2 \pi}{L a_{\text{lat}}} (m_x' + \frac{\vartheta_x}{2\pi}, m_y' + \frac{\vartheta_y}{2\pi}, m_z' + \frac{\vartheta_z}{2\pi}),
\end{equation}
then integrate over the common shift $(\vartheta_x, \vartheta_y, \vartheta_z) \in [0, 2 \pi)^3 \equiv T^3$ ($T^3$ stands for the three-dimensional torus).

In order to calculate $J_0 + \sum_{i\ge 1}z_i J_i$ we rewrite (\ref{eq: ps}) with $L = 1$ as
\begin{equation}
  \sum_{j} \exp\left[\mbox{i}(\kappa - \kappa') (R_i - R_j)\right] = \left(\frac{2\pi}{a_{\text{lat}}}\right)^3   \sum_{m_x, m_y, m_z}
  \delta\left(\kappa - \kappa' + \frac{2\pi}{a_{\text{lat}}} (m_x, m_y, m_z)\right).
\end{equation}
Since $2\pi/a_{\text{lat}}$ is the lattice constant of the reciprocal lattice, only the term with $m_x = m_y = m_z = 0$
remains in the last sum, and the Dirac delta suppresses one integration. Therefore, $J_0 + \sum_{i\ge 1} z_i J_i$ and, thus, $\tilde{\chi}(0)$
 can be computed as a single integral over the BZ.
\end{widetext}

In insulators, the integration over $\vartheta$'s is approximated by a sum over an equally spaced grid (the trapezoids method).
The number of required $\vartheta$-points in each Cartesian direction gets smaller as the larger supercells are considered.
Ultimately, just one $\vartheta$-point is sufficient; it can be chosen as, e.g., $\vartheta_x = \vartheta_y = \vartheta_z = 0$ or $\vartheta_x = \vartheta_y = \vartheta_z = \pi$,
in correspondence with the boundary conditions imposed on the electronic wavefunctions. In contrast, in semimetals, if the boundary conditions
dictate $\vartheta = 0$, finite summation is not appropriate because the denominator $E_k - E_{k'}$ in Eq.\,7  renders the quantity undefined.
Therefore, a special set of $\vartheta$-points has been chosen here, which is equivalent to the transformation $\vartheta_i' \mapsto \vartheta_i = \vartheta_i' - \sin\vartheta_i'$
under the BZ integral. This transformation, besides possessing analytic properties, preserves periodicity while $d \vartheta_i/d \vartheta_i' = 0$ at $\vartheta_i' = 0$.
As a result, the divergence of the integral at the origin cancels with the zero of the Jacobian of the transformation and the integral can be computed
with the trapezoids method. Indeed, the singularity of the integrand at $\vartheta = 0$ is integrable, as long as the dimensionality is sufficient
and the band structure is well-behaved ($k$-linear terms, $k^3$-terms, and the anisotropy may play a role here).
The RKKY (i.e., \ intraband) term is omitted in this discussion, as the density of states vanishes if the Fermi energy $E_F \to 0$.

The computations have been performed with an efficient algorithm based on the fast Fourier transform on a $16 \times 16 \times 16$ supercell (16384 cation lattice sites) that also
determines  the grid density of $k$ and $k'$ points employing periodic boundary conditions.
For Cd$_{1-x}$Mn$_x$Te and Hg$_{1-x}$Mn$_x$Te, the grids with up to two and eight different $\vartheta$ values have been employed, respectively.
Figure \ref{fig: cvg} demonstrates that the magnitude of $\Theta_{he}$ in Hg$_{1-x}$Mn$_x$Te converges with the number of employed $\vartheta$ values. This means that
in contrast to the static dielectric function \cite{Liu:1969_PRL,Broerman:1970_PRL}, the spin susceptibility, though enhanced, is not singular at $q \to 0$ in the symmetry-induced
zero-gap semiconductors.  Aitken's delta-squared process served to accelerate the integration convergence. We have checked that $\Theta_0$ calculated by the single integral, as outlined in a preceding paragraph, and after subtracting $J_0$, is in excellent numerical agreement (better than 1\,K) with the value obtained by summing  $z_iJ_i$, $i\ge 1$, obtained by the double BZ integration. Actually, the coupling to the nearest neighbors  contributes over 50\% to the value of $\Theta_0$.

over 50

\begin{table}[tb]
\caption{Consecutive nearest-neighbor exchange energies $-(J_i = J_{i,hh} + J_{i,ee} + J_{i,he})$, $i = 1, 2, 3, 4$ (from the fourth-order perturbation theory) and Curie-Weiss parameter $-\Theta_0$ in Kelvins compared to experimental results. Contributions from the superexchange ($hh$), electron-electron ($ee$), and interband  ($he$) terms to $\Theta_0$ are also shown.}
\begin{tabular}{c||c|c||c|c}
\hline
\hline
 \multicolumn{1}{c||}{} &  \multicolumn{2}{c||}{Cd$_{1-x}$Mn$_x$Te}&  \multicolumn{2}{c}{Hg$_{1-x}$Mn$_x$Te} \\
\hline
&theory&expl&theory&expl\\
\hline
\hline
$-J_1$& $ 9.77$&$ 6.3 \pm 0.3$ \cite{Larson:1986_PRB}&$6.46$&$5.1 \pm 0.5$ \cite{Galazka:1988_JMMM}\\
              && $6.15 \pm 0.05$ \cite{Cherbunin:2020_PRB} &&$4.3 \pm 0.5$ \cite{Lascaray:1989_PRB}\\
              \hline
$-J_2$& $ 0.810$&$ 1.9 \pm 1.1$ \cite{Larson:1986_PRB} & $0.842$&\\
      &         &$1.80 \pm 0.05$ \cite{Cherbunin:2020_PRB} &&\\
      \hline
$-J_3$& $ 0.352$&$0.4 \pm 0.3$ \cite{Larson:1986_PRB}  & $0.394$&\\
     &          &$ 1.39 \pm 0.05$ \cite{Cherbunin:2020_PRB}&    &\\
     \hline
$-J_4$& $ 0.255$&$0.81 \pm 0.05$ \cite{Cherbunin:2020_PRB} & $0.467$& \\
\hline
$-\Theta_{0}$& $801$&$470 \pm 34$ \cite{Spalek:1986_PRB}& $666$&$500 \pm 10$\cite{Spalek:1986_PRB}\\
                                                          &&&&$660 \pm 88$\cite{Lewicki:1988_PRB}\\
\hline
\hline
\end{tabular}
\begin{tabular}{c|c|c|c||c|l c}
 \multicolumn{5}{c}{theory}  \\
\hline\hline
$-J_{1,he}$& $-0.396$&$-\Theta_{hh}$& $772$ &$-1.583$&$651$&\phantom{two}\\
$-J_{2,he}$&$0.421$&$-\Theta_{ee}$& $8.9$ &$0.492$&$19.0$&\phantom{two}\\
$-J_{3,he}$&$0.131$&$-\Theta_{he}$& $20.0$ &$0.112$&$-3.9$&\phantom{two}\\
$-J_{4,he}$&$0.061$& &&$0.154$&&\phantom{two}\\
\hline
\hline
\end{tabular}
\label{tab:J_Theta}
\end{table}

\begin{figure*}[ht!]
 \includegraphics[scale=0.7]{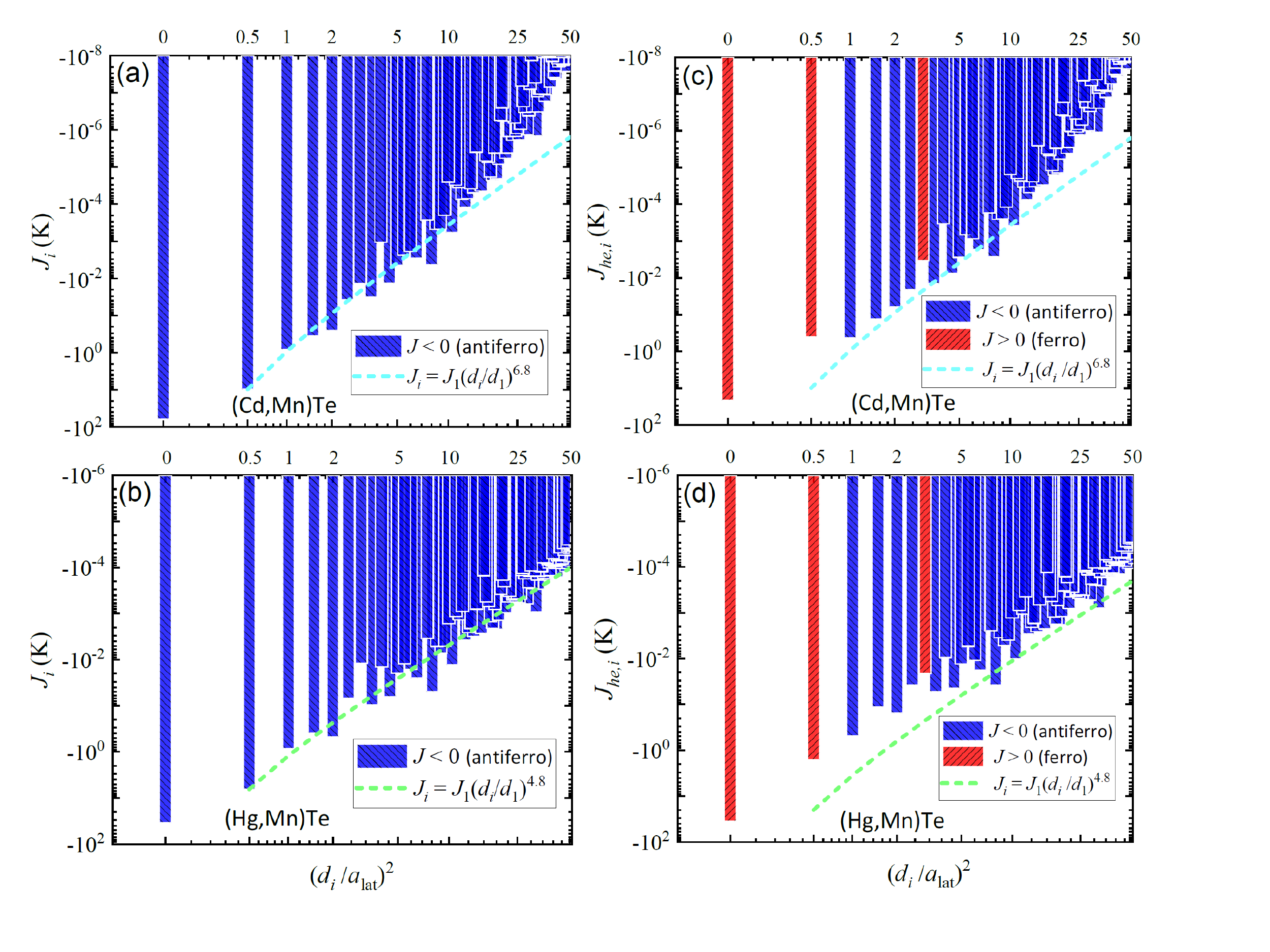}
  \caption{Computed total exchange energies $J_i$ (a,b) and the interband BR contribution $J_{i,he}$ (c,d) for Mn pairs (including the self-interaction values $J_0$)
vs.\,Mn-Mn distances $d_i$ in the unit of the lattice parameter $a_{\text{lat}}$ for CdTe (a,c) and HgTe (b,d).  Dashed lines indicate
$J_i \sim d_i^{-n}$ with $n = 6.8$ and $4.8$, as found experimentally for Cd$_{1-x}$Mn$_x$Te \cite{Twardowski:1987_PRB}
and Hg$_{1-x}$Mn$_x$Te \cite{Galazka:1995_JMMM}, respectively. }
\label{fig2}
\end{figure*}

{\em Discussion of theoretical results vis-\`a-vis experimental data}. As expected theoretically for the random distribution of magnetic ions, experimental values of $\Theta_{\text{CW}}$ show linear dependence on $x$ in II-VI DMSs \cite{Spalek:1986_PRB,Sawicki:2013_PRB}. As shown in Table I, our theory and together with the employed tight-binding parametrization explains the interaction sign but overestimate by 60\% the absolute values of $J_1$ and $\Theta_0$ in Cd$_{1-x}$Mn$_x$Te and by 30\% in the case of Hg$_{1-x}$Mn$_x$Te.

 Extensive experimental studies of spin-glass freezing temperature $T_f$ in wide-gap Mn- and Co-based DMSs, including Cd$_{1-x}$Mn$_x$Te, indicate that $\Theta_{\text{CW}} \gg T_f \sim x^{\alpha}$, where $\alpha = 2.25\pm 0.1$ \cite{Galazka:1980_PRB,Twardowski:1987_PRB,Galazka:1995_JMMM,Stefanowicz:2013_PRB}. A scaling argument \cite{Twardowski:1987_PRB,Smith:1975_JPF} then implies $J_i \sim d_i^{-n}$, where $d_i$ is the distance between spin pairs and $n = 3\alpha =6.8 \pm 0.3$ \cite{Twardowski:1987_PRB}.
Figure\,\ref{fig2}(a) demonstrates that the  dependence of $J_i$ on $d_i$ obtained here for Cd$_{1-x}$Mn$_x$Te is in agreement with the experimentally determined power law. However, an exponential decay would describe the computed data over a wider range of $d_i$.

Comparing $J_{i,he}$ values displayed in Figs.\,\ref{fig2}(c,d) to $J_l$ data in  Figs.\,\ref{fig2}(a,b), we find that the BR mechanism dominates at large $d$. It decays exponentially with $d$ in the wide-gap Cd$_{1-x}$Mn$_x$Te but for topological zero-gap Hg$_{1-x}$Mn$_x$Te, $J_{i,he}(d)$  shows a power-law dependence, also at large $d$. This behavior accounts for a relatively weak decay of $T_f$ with decreasing $x$ in Hg$_{1-x}$Mn$_x$Te \cite{Mycielski:1984_SSC}, leading to $n = 4.8$ \cite{Galazka:1995_JMMM}. As seen in Fig.\,\ref{fig2}(b), this value of $n$ is consistent with our theoretical results, though we  have to note that a considerable shift of bands with $x$ is expected in topological materials, while our computations have been performed for Mn pairs in HgTe.
As shown in Table~\ref{tab:J_Theta}, the relevant ferromagnetic and antiferromagnetic contributions $J_{i,he},\,i\ge 1$ actually cancel each other in $\Theta_{he}$, making the contribution of the BR mechanism to $\Theta_{0}$ negligible in both compounds. This explains why no effect of gap opening on $\Theta_{\text{CW}}$ was found in Hg$_{1-x-y}$Cd$_{y}$Mn$_x$Te \cite{Lewicki:1988_PRB}. At the same time, it is clear from Fig.\,\ref{fig: cvg} that the inclusion of the self-interaction term $J_{0,he}$, would drastically increase the magnitude of $\Theta_{he}$.

{\em Conclusions and outlook.} Our results demonstrate that the interband BR term changes the sign from ferromagnetic to antiferromagnetic as a function of Mn pair distance, with the behavior contradicting the Van Vleck-like approach that predicts only the ferromagnetic coupling \cite{Yu:2010_S}. Such an alternating sign, reflecting the presence of both ferromagnetic and antiferromagnetic excitations in $\tilde{\chi}(q)$ \cite{Bednik:2020_PRB}, significantly reduces the role of the interband contribution making the superexchange to determine whether a spin-glass or a ferromagnet becomes the magnetic ground state, the case of Mn$^{2+}$ in II-VI compounds and Mn$^{3+}$ in GaN, respectively.

There are persisting uncertainties concerning the distribution (random vs.~clustering \cite{Chang:2014_PRL}) and the location of transition metal (TM) impurities in the tetradymite lattice (substitutional vs.~interstitial positions in the van der Waals gap \cite{Ruzicka_2015_NJP}). Nevertheless, a series of arguments allows extending the conclusion about the dominance of the superexchange to topological tetradymite chalcogenides doped by substitutional V, Cr, or Fe ions, whose magnetism has so far been merely attributed to the Van Vleck mechanism \cite{Ke:2018_ARCMP,Tokura:2019_NRP}. (i) These impurities appear isoelectronic \cite{Ke:2018_ARCMP,Tokura:2019_NRP,Satake:2020_PRM}, which means that $d$ orbitals remain fully occupied or empty.  Moreover, as in other DMSs,  correlations, together with the Jahn-Teller effect and dilution, enhance the $d$ orbital localization further on.  Accordingly, the double-exchange scenario, put forward when interpreting {\em ab initio} results  \cite{Vergniory:2014_PRB,Peixoto:2020_QM}, is not valid. (ii) Another {\em ab initio} study reveals the insensitivity of the spin-spin coupling energy of the band inversion \cite{Kim:2017_PRB}, the finding contradicting the Van Vleck model. (iii) Recent  studies of x-ray magnetic circular dichroism and resonant photoelectron spectroscopy demonstrate similarities of $p$-$d$ hybridization effects in V- or Cr-doped (Bi$_x$Sb$_{1-x}$)$_2$Te$_3$ \cite{Peixoto:2020_QM} and II-VI DMSs \cite{Kacman:2001_SST}, in particular, stronger hybridization of $t_{2g}$ TM levels compared to the $e_g$ case, which implies a similar physics of spin-spin coupling as found in tetrahedrally coordinated DMSs.  (iv) As superexchange prevails over the interband Van Vleck mechanism in the zero-gap case,  it should dominate even more strongly in the gapped topological systems. (v) From the direct computations for tetrahedral systems carried out here and previously \cite{Blinowski:1996_PRB,Simserides:2014_EPJ,Blinowski:1995_MSF}, supported by experimental data \cite{Stefanowicz:2013_PRB,Twardowski:1987_PRB,Galazka:1995_JMMM,Twardowski:1990_JAP,Watanabe:2019_APL}, we know that the superexchange is ferromagnetic for $d^3$ and $d^4$, whereas it is antiferromagnetic for $d^5$ and $d^6$ cases. According to experimental results \cite{Ke:2018_ARCMP,Tokura:2019_NRP,Satake:2020_PRM} and {\em ab initio} studies \cite{Vergniory:2014_PRB}, the same sequence occurs in tetradymite topological insulators, except for the Mn case, as Mn acts as an acceptor \cite{Hor:2010_PRB}, so that the RKKY interaction accompanies the antiferromagnetic superexchange, such as (Ga,Mn)As \cite{Dietl:2014_RMP,Sato:2010_RMP}.   Altogether, these arguments indicate that the TM charge state and coordination, more than a topological class, govern the magnetic properties of DMSs.

{\em Acknowledgments.}
The work is supported by the Foundation for Polish Science through the International Research Agendas program co-financed by the
European Union within the Smart Growth Operational Programme. One of us (JAM) acknowledges support of the National Science Centre (Poland) under the grant UMO-218/31/BST3/03758.
We acknowledge the access to the computing facilities of the Interdisciplinary
Center of Modeling at the University of Warsaw, Grant No.~G68-12.
\clearpage
\begin{center}
\textbf{\Large Supplemental Material} \\
\end{center}%
\renewcommand{\thefigure}{S\arabic{figure}}
\renewcommand{\thetable}{S\arabic{table}}
\renewcommand{\theequation}{S\arabic{equation}}
\setcounter{section}{0}
\setcounter{equation}{0}
\setcounter{figure}{0}
\setcounter{table}{0}
\begin{widetext}
\section{Overview of various exchange mechanisms}
Since the exchange part of the Coulomb energy decays exponentially with the distance between magnetic ions, the spin-spin interaction in transition-metal compounds is dominated by indirect coupling {\em via} non-magnetic states residing in-between localized spins \cite{White:2007_B,Kacman:2001_SST}. Quite generally, the presence of $d$ orbitals brought about by transition metals leads to additional delocalization of valence electrons, i.e., to the lowering of their quantum mechanical kinetic energy, as illustrated in Fig.~\ref{figS1} \cite{Bonanni:2010_CSR}. The effect is spin dependent as, according to the Hund's rule and due to the on-site Hubbard energy $U$, $d$-orbitals are spin-polarized in transition metals.

\begin{figure}[b!]
 \includegraphics[scale=0.6]{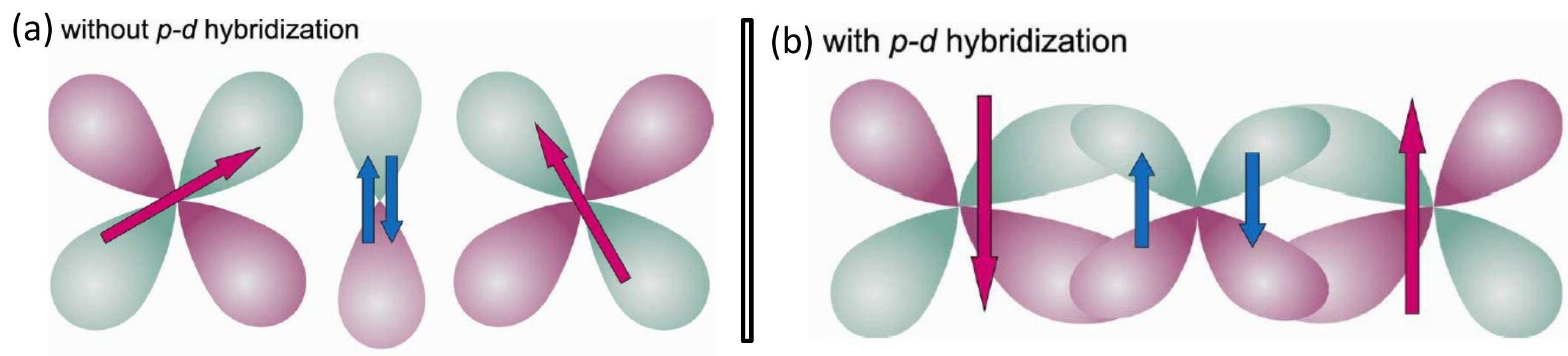}
  \caption{Schematic illustration of open $d^5$ orbitals at the transition metal cations and entirely occupied $p$ orbitals at the anion without (a) and
  with (b) $p$-$d$ hybridization.  A spin-dependent
shift of the orbitals (described by the fourth order perturbation theory), occurring for an antiferromagnetic arrangement of
Mn spins (red arrows) shown in (b), leads to lowering of the
quantum kinetic energy associated with delocalization of electrons over the $p$ and $d$ orbitals. Note that virtual $p$-$d$ transitions occurs only for
antiparallel orientation of Mn (red arrows) and anion spins (blue arrows).}
\label{figS1}
\end{figure}

Recently, we proposed  a formal theory of the indirect spin-dependent interaction resulting from hybridization between $d$-orbitals of magnetic ions and $sp$ band states \cite{Sliwa:2018_PRB}.  That theory, within the second order in the hybridization energy $V_{\text{hyb}}$ provides exchange integrals describing Kondo-like coupling between localized spins and band states $J_{sp-d}$ \cite{Kacman:2001_SST,Sliwa:2018_PRB,Autieri:2021_PRB}, whereas the fourth order in $V_{\text{hyb}}$ allows to obtain the exchange tensor of the interaction between pairs of the localized spins \cite{Kacman:2001_SST,Sliwa:2018_PRB,Larson:1988_PRB}. This procedure encompasses on equal footing various spin-spin exchange mechanisms, including superexchange, Ruderman-Kittel-Kasuya-Yosida (RKKY) and Bloembergen-Rowland-Van Vleck (BR-VV) interactions, the latter two usually described within the second order perturbation theory in $J_{sp-d}$, as discussed in the next section. We recall  that these two mechanisms  of coupling between localized spins can be viewed as mediated by spin-polarization of band,
carriers and/or occupied bands, in an analogy to intraband and interband dielectric polarizations. However, in the absence of band carriers, the dominant interaction is the superexchange described in the fourth order in $V_{\text{hyb}}$ by a diagram depicted in Fig.~S2(a) \cite{Kacman:2001_SST}. As shown in the main body of our paper, if the self-interaction term is treated correctly, i.e., omitted when considering pairwise interactions, the superexchange [Fig.~S2(a)]
dominates over the interband BR-VV contribution [Fig.~S2(b)], even in the case of zero-gap Hg$_{1-x}$Mn$_x$Te.

\begin{figure}[t!]
 \includegraphics[scale=0.7]{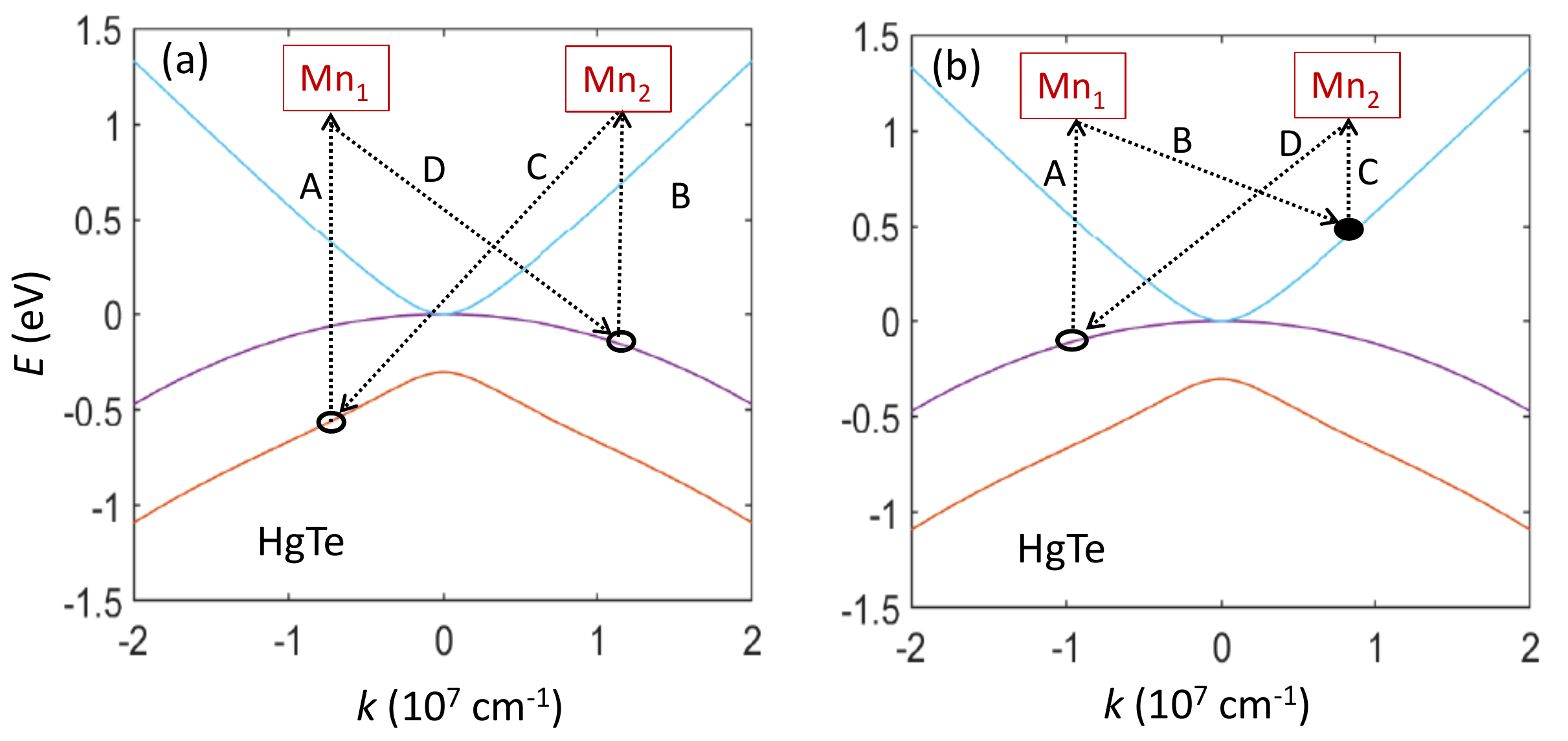}
  \caption{Examples of four virtual transitions allowed by  hybridization between band states and $d$ orbitals of two Mn ions (labeled 1 and 2) in zero-gap HgTe. These processes account for (a) superexchange (the $hh$ contribution) and (b) Bloembergen-Rowland-Van Vleck coupling (the $eh$ contribution). }
\label{figS2}
\end{figure}

\section{Fourth-order vs. second-order perturbation theory}

Typically, as mentioned in the previous Section, the RKKY interaction is assumed to originate from a Kondo-like contact interaction $\hat {\cal{H}}_{pd}$,
\begin{equation}
\hat {\cal{H}}_{pd} =  -\frac{1}{\mathcal{V}} \, \left(\hat{J}_{pd}(\mathbf{r}) \hat{\mathbf{s}}\right) \cdot \hat{\mathbf{S}}_i,
\end{equation}
where $\hat{\mathbf{s}}$ and $\hat{\mathbf{S}}_i$ are the band electron's and impurity's spin
operators, respectively, whereas the usual exchange integral $J_{pd}$ has been
replaced with a Hermitian operator $\hat{J}_{pd}(\hat{\mathbf{r}})$. For an Anderson impurity, $\hat {\cal{H}}_{pd}$ is considered
a second-order effective Hamiltonian that can be obtained via the Schrieffer-Wolff transformation \cite{Schrieffer:1966_PR}.
We imagine that this operator is a product,
$\hat J_{pd}(\hat{\mathbf{r}}) \equiv J_{pd} \hat P(\hat{\mathbf{r}})$, of a constant $J_{pd}$ and a Hermitian operator $\hat P(\hat{\mathbf{r}})$,
the latter restricting the interaction to the tight-binding orbitals of appropriate symmetry
[namely, the $p$ orbitals of the anion; in the effective-mass approximation $\hat P$ includes
a Dirac delta at the impurity position $\mathbf{R}_i$, $\delta(\hat{\mathbf{r}} - \mathbf{R}_i)$].
The Hamiltonian for two impurities $(i, j)$ reads
\begin{equation}
  \hat {\cal{H}}_{\text{eff}(2)} = -\frac{1}{\mathcal{V}} \,
    \hat J_{pd}(\hat{\mathbf{r}}) \hat{\mathbf{s}} \cdot \left( \hat{\mathbf{S}}_i + \hat{\mathbf{S}}_j \right).
\end{equation}
By applying a second order thermodynamic perturbation theory we obtain the Landau free energy $\Omega$
for the gas of band electrons,
\begin{equation}
  \Omega^{(2 \times 2)} = \sum_{k,k'} \frac{f(E_k)}{E_k - E_{k'}} \left< k \middle| \hat H_{\text{eff}(2)} \middle| k' \right>
    \left< k' \middle| \hat H_{\text{eff}(2)} \middle| k \right>
\end{equation}
($2 \times 2$ stands for the second order perturbation theory on the second order effective Hamiltonian, yielding effectively the fourth order).
It includes an interaction term,
\begin{equation}
  \Omega^{(2 \times 2)}_{i,j} = \sum_{\alpha\beta} \frac{2 S_{i\alpha} S_{j\beta}}{\mathcal{V}^2} \sum_{k,k'}
    \frac{f(E_k)}{E_k - E_{k'}} \left< k' \middle| \hat J_{pd}(\mathbf{r}) \hat s_\alpha \middle| k \right>
    \left< k \middle| \hat{J}_{pd}(\hat{\mathbf{r}}) \hat{s}_\beta \middle| k' \right>,
\end{equation}
where the impurities are assumed to be in spin-coherent states, $\left< \hat S_{i\alpha} \right> = S_{i\alpha}$.
Therefore the low-temperature interaction constants, $J^{(2 \times 2)}_{ij,\alpha\beta}$, are given by
\begin{equation}
  J^{(2 \times 2)}_{ij,\alpha\beta} = -\frac{J_{pd}^2}{\mathcal{V}^2} \sum_{k,k'}
    \frac{f(E_k)}{E_k - E_{k'}} 
\left< k' \middle| \hat P \hat s_\alpha \middle| k \right>
    \left< k \middle| \hat P \hat s_\beta \middle| k' \right>.
\end{equation}
Since (according to Schrieffer and Wolff \cite{Schrieffer:1966_PR})
\begin{equation}
  J_{pd} = \frac{1}{S} \frac{U}{(E_d - E_F) (E_d + U - E_F)} \left| \tilde V_{k_F} \right|^2,
\end{equation}
we obtain
\begin{equation}
  J^{(2 \times 2)}_{ij,\alpha\beta} = -\frac{1}{2 (2 S)^2} \frac{U^2}{(E_d - E_F)^2 (E_d + U - E_F)^2} \sum_{k,k'}
    \frac{f(E_k) - f(E_{k'})}{E_k - E_{k'}} 
\left< k' \middle| \hat Q \hat \sigma_\alpha \middle| k \right>
    \left< k \middle| \hat Q \hat \sigma_\beta \middle| k' \right>, \qquad
\end{equation}
with $\hat Q = \frac{1}{\mathcal{V}} \left| \tilde V_{k_F} \right|^2 \hat P$.
Furthermore, by writing
\begin{equation}
  \hat Q = \hat V_{\mathrm{hyb}}^{\dagger} \left( \sum_{m; a=\uparrow,\downarrow} \left| d_i m a \right> \left< d_i m a \right| \right) \hat V_{\mathrm{hyb}}
\end{equation}
and assuming that the hybridization is spin-independent:
\begin{equation}
  \hat Q \hat \sigma_\alpha = \hat V_{\mathrm{hyb}}^{\dagger}
    \left( \sum_{m; a,b=\uparrow,\downarrow} \left| d_i m a \right> \left< a \middle| \sigma_\alpha \middle| b \right> \left< d_i m b \right| \right) \hat V_{\mathrm{hyb}},
\end{equation}
we reproduce some of the $he$ terms (those with the singular denominator), neglecting the dependence on the band energies $(E_k, E_{k'})$.

The remaining $he$ terms (those proportional to $2/U$) can be derived from the second-order quasi-degenerate perturbation theory \cite{Winkler:2003_B} as follows.
Consider the six-dimensional Hilbert space for two Anderson impurities occupied by two electrons. The ground state corresponds to a single occupation
of each impurity and is four-fold spin-degenerate, the remaining two states are higher in energy by $U$. The $2/U$ terms appear on restricting the Hamiltonian
to the four-dimensional ground state via application of the Winkler's perturbation theory (in the second order).

\section{Bloembergen-Rowland mechanism: integration convergence}
\label{sec: brpar}

In this Section, the BR-VV interaction in a zero-gap semiconductor ($E_g = 0$) with spherical, parabolic bands is considered in the case of accidental degeneracy, $E_{c(v)} = +(-)\hbar^2k^2/2m_{c(v)}$.
The double reciprocal-space integral that arises is not absolutely convergent. It may be truncated at a finite momentum \cite{Bloembergen:1955_PR,Lee:1988_PRB}
or understood in the improper sense \cite{Bastard:1979_PRB,Lewiner:1980_JPC}, i.e., as the limit $\lim_{k_{\text{max}} \to {+\infty}} \int_{k = 0}^{k = k_{\text{max}}}$.
It turns out that the order of taking the limit in the two integrals (if the integral is calculated as an iterated one) may affect the result
of the double-integration procedure, as evidenced by the violation of the expected symmetry between the valence and conduction bands ($m_c \leftrightarrow m_v$) \cite{Bastard:1979_PRB}.
In the extreme case, even the condition $J_0 + \sum_{i\ge 1} z_i J_i > 0$,  may be violated. This difficulties appear as unavoidable and make estimation of the interaction
within a simple model rather complex, as elaborated previously. With the improper sense in place and the Fermi level in the conduction band ($\hbar k_F = \sqrt{2 m_c E_F}$),
the Hamiltonian for a system of two localized spins assumes the form:
\begin{eqnarray}
  \lefteqn{{\cal{H}}(R_{ij}) = - \frac{2 J_{cv}^2}{(2\pi)^3 \hbar^2} \frac{(m_c m_v)^{3/2}}{(m_c+m_v)^2}\frac{\exp(-\sqrt{m_v/m_c} k_F R_{ij})}{R_{ij}^4} \times {}} \nonumber \\
  & & {} \times \Biggl[ \left( 2 + \frac{m_c + m_v}{\sqrt{m_c m_v}} k_F R_{ij} \right) \cos(k_F R_{ij}) + \left( \frac{m_v - m_c}{\sqrt{m_c m_v}} + \frac{m_c + m_v}{m_c} k_F R_{ij} \right) \sin(k_F R_{ij}) \Biggr] \mathbf{S}_i \cdot \mathbf{S}_j,
\end{eqnarray}
where $J_{cv}$ is an interband $sp$-$d$ exchange integral. In contrast to Ref.\,\onlinecite{Bloembergen:1955_PR}, where a hard momentum cut-off $k_t$ was introduced, the expression is not integrable at $R_{ij} = 0$, and as such,
is significantly different from Eq.\,32 in Ref.\,\onlinecite{Bloembergen:1955_PR}. Assuming a real-space cut-off at $R_{\text{min}}$,
\begin{eqnarray}
  \lefteqn{ \Theta_{he} = N_0 \left( \frac{J_{cv}}{\pi \hbar} \right)^2 \frac{m_c m_v}{(m_c + m_v)^2} \frac{\exp(-\sqrt{m_v/m_c} k_F R_{\text{min}})}{R_{\text{min}}} \times {}} \nonumber \\
  & & {} \times  \left[ 2 \sqrt{m_c m_v} \cos(k_F R_{\text{min}}) - (m_c - m_v) \sin(k_F R_{\text{min}}) \right] \frac{S (S + 1)}{3}.
\end{eqnarray}
The oscillations as a function of $k_F R_{\text{min}}$ are damped exponentially with $\alpha = \sqrt{m_v/m_c}$
and shifted in phase by $\phi = \arctan[(m_c - m_v)/2\sqrt{m_c m_v}]$. Taking, for instance,  $R_{\text{min}} = (4 \pi N_0)^{-1/3}$,
and parameters for intrinsic HgTe, $k_F =0$, $a_{\text{lat}} = 0.646$\,nm, $N_0J_{cv} =-0.6$\,eV; $m_c = 0.3m_0$, and $m_v = 0.45m_0$, we obtain ferromagnetic $\Theta_{he} = 85$\,K.

An alternative procedure is proposed here which amounts to regularizing the momentum integrals by introducing under the integral an additional weight,
$dk \to \exp(-ak) \, dk$, $a \to 0^{+}$. One proceeds with integration over one of the momenta ($k$) and takes the limit $a \to 0^{+}$. Then,
an additional factor appears which exponentially divergent and must be suppressed by $dk' \to \exp(-a'k') \, dk'$ with sufficiently large $a'$.
However, the result of the second integration is analytic with respect to $a'$ and can be continued to $a' = 0^{+}$, where its value vanishes as $a' \to 0^{+}$.
Furthermore, although the exchange constant diverges as $1/a'$ if integrated over the real space ($R_{ij}$), the real space integral vanishes again
if the vicinity of $R_{ij} = 0$ is omitted from the integration. One concludes that this simple model predicts $\Theta_{he} = 0$, and the interaction
vanishes except for $R_{ij} \to 0$. In practice, values of $a$ and $a'$ of the order of the lattice constant are appropriate, and result
in a real-space-distance dependence of the exchange constant similar to that presented in Fig.\,2 in the main text.
\end{widetext}


%

\end{document}